\documentclass{article}
\usepackage{dcase2020,graphicx,url,times,booktabs,tabularx}
\usepackage{amsmath,amssymb,amsfonts}
\usepackage{subfig}
\usepackage[export]{adjustbox}
\usepackage[linesnumbered,ruled,vlined]{algorithm2e}
\title{Conditioned Time-Dilated Convolutions for Sound Event Detection}
\name{Konstantinos Drossos$^{1}\sthanks{K. Drossos and T. Virtanen wish to acknowledge CSC-IT Center for Science, Finland, for computational resources. Part of the computations leading to these results was performed on a TITAN-X GPU donated by NVIDIA to K. Drossos.}$,
      Stylianos I. Mimilakis$^{2}\sthanks{Stylianos I. Mimilakis is supported in part by the German Research Foundation (AB 675/2-1, MU 2686/11-1).}$,
      Tuomas Virtanen$^{1*}$}
\address{$^1$ Audio Research Group, Tampere University, Tampere, Finland\\
         \{firstname.lastname\}@tuni.fi\\
        $^2$ Fraunhofer-IDMT, Ilmenau, Germany\\mis@idmt.fraunhofer.de}
\begin{document}
\ninept
\maketitle
\begin{sloppy}
\begin{abstract}
Sound event detection (SED) is the task of identifying sound events along with their onset and offset times. A recent, convolutional neural networks based SED method, proposed the usage of depthwise separable (DWS) and time-dilated convolutions. DWS and time-dilated convolutions yielded state-of-the-art results for SED, with considerable small amount of parameters. In this work we propose the expansion of the time-dilated convolutions, by conditioning them with jointly learned embeddings of the SED predictions by the SED classifier. We present a novel algorithm for the conditioning of the time-dilated convolutions which functions similarly to language modelling, and enhances the performance of the these convolutions. We employ the freely available TUT-SED Synthetic dataset, and we assess the performance of our method using the average per-frame $\text{F}_{1}$ score and average per-frame error rate, over the 10 experiments. We achieve an increase of 2\% (from 0.63 to 0.65) at the average $\text{F}_{1}$ score (the higher the better) and a decrease of 3\% (from 0.50 to 0.47) at the error rate (the lower the better). 
\end{abstract}


\section{Introduction}\label{sec:intro}
Sound event detection (SED) is a typical machine listening task, and is the identification of activities of different sound events (e.g. ``people speaking'', ``dog barking'') together with the detection of their onset and offset times. This is usually implemented by identifying the active sound events in a sequence of short-time windows of the recording (e.g. 0.02 seconds), effectively detecting when these sound events are starting and ending~\cite{li:2020:icassp,drossos:2020:ijcnn}. Typical SED methods are usually based on deep neural networks (DNNs), and consist of a learnable feature extractor and a module for detecting temporal patters on top of the output of the feature extractor~\cite{kapka:2019:dcase,grondin:2019:dcase}. Typically in the literature, the feature extractor is implemented using convolutional neural networks (CNNs), and the identification of the temporal patterns using recurrent neural networks (RNNs). Such set-up is seen in the wide-spread convolutional-recurrent neural networks (CRNN) architecture~\cite{cakir:2017:taslp} and its adoption to different SED and audio signal processing tasks and methods, e.g.~\cite{malik:2017:smc,adavanne:2019:jstsp,adavanne:2017:eusipco,cakir:2017:eusipco,adavanne:2017:icassp}. 

Recently, there is an interest in re-examining the usage of the CNNs and RNNs in SED and audio classification, in general. There are published papers focusing on the tasks of sound event tagging and SED, where the CNNs are been replaced by depthwise separable (DWS) convolutions~\cite{drossos:2020:ijcnn,fonseca:2019:dcase}. DWS convolutions are a factorized version of the typical convolutions in the CNNs, by first learning the spatial information and then processing the cross-channel patterns~\cite{sifre:2014:phd,guo:2018:bmvc}. This factorization of the convolution carried out by typical CNNs, leads to considerable fewer trainable parameters and smaller memory footprint of the CNNs, reducing the computational complexity by a factor of $K^{-1}_{\text{o}}+(K_{\text{h}}\cdot K_{\text{w}})^{-1}$, where $K_{\text{h}}$ and $K_{\text{w}}$ are the height and width of the CNN kernel, and $K_{\text{o}}$ are the output channels of the CNN~\cite{drossos:2020:ijcnn}. This reduction of the computational complexity, renders the SED and sound event tagging systems more fit for deployment in devices with restricted resources, e.g. embedded and wearable devices. 

The employment of the DWS convolutions has not only yielded DNNs with smaller memory footprint, but in some cases also boosted the performance of the DNN~\cite{drossos:2020:ijcnn,fonseca:2019:dcase,howard:2017:arxiv}. For example, YAMNET\footnote{\url{https://github.com/tensorflow/models/tree/master/research/audioset/yamnet}} is a sound event tagging system, using specifically DWS convolutions and focusing on having a small amount of total parameters. In~\cite{fonseca:2019:dcase} is presented another system for sound event tagging, implemented with 13 layers of DWS convolutions. Though, nor the system in~\cite{fonseca:2019:dcase}, neither YAMNET, have implemented any explicit functionality for detecting long temporal patterns. Such long temporal patters and apparent for different classes or between different classes. For example, ` ``car passing by'' is more likely to appear before or after a ``car horn'', and a ``water tap running'' is an event with longer duration that water ``flushing'' or ``splashing''. 

Dilated convolutions have been proposed as way to increase the capabilities of learning long temporal patterns for CNNs~\cite{yu:2016:mca}. In a nutshell, the kernel of a dilated convolution is expanded, having distance between its elements. This allows the kernel of the dilated convolution to be applied on every other $N$ elements (where $N$ is the dilation factor) of the input patch to the kernel, increasing the receptive field of the kernel without increasing its parameters~\cite{yu:2016:mca,holschneider:1990:springer,shensa:1992:tsp}. Dilation of the kernel can be applied to any combination of its dimensions (e.g. dilation only at the dimension of time or only at the dimension of features) or all of them together. 

Recently, dilated convolutions have been employed for SED, showing promising results. Specifically,~\cite{li:2020:icassp} presented an approach where dilated convolutions were used together with RNNs for the task of SED, focusing explicitly on the exploitation and learning of long temporal patterns.~\cite{drossos:2020:ijcnn} presented an adaptation of the CRNN architecture, using DWS and dilated convolutions, using dilation only in the dimension of time (i.e. time-dilated convolution). With the employment of the DWS and time-dilated convolutions, the system in~\cite{drossos:2020:ijcnn} had 85\% less parameters than the CRNN, while at the same time, achieved better performance on a typical SED dataset. The improvement achieved with the dilated convolutions indicates that these convolutions can be used for the effective modelling of longer temporal context, similarly to what RNNs do. 

A typical process that is applied to models dealing with sequences and using an RNN and a classifier, is the conditioning of each input to the RNN with the immediate previous prediction of the classifier. For example, in an RNN-based sequence-to-sequence model, the input to the first RNN of the decoder can be conditioned with the predictions of the classifier of the model. This conditioning is called language modelling (LM), and has been incorporated extensively in natural language modelling~\cite{sutskever:2014:neurips,auli:2013:emnlp,bengio:2015:neurips} and, recently, in SED~\cite{drossos:2019:dcase}. The benefit of LM is that the RNN can be informed on the previous predictions of the classifier, enforcing a contextual modelling of the class activities by the RNN, and leading to the learning of longer intra- and inter-class temporal patterns. For example, specific sound events tend to appear sequentially in time, e.g. ``car horn'' and ``car passing by''. Other sound events can be active in non-consecutive short time frames (e.g. around 40 ms), for example ``footsteps''. Using LM for SED, a method can benefit by modelling these class dependencies, resulting in increased performance~\cite{drossos:2019:dcase}. 

In this work we present a novel algorithm and a method for employing a mechanism similar to LM, but for time-dilated convolutions (i.e. dilated convolutions using dilation only at the time dimension). We build upon previous methods for SED, which employ LM~\cite{drossos:2019:dcase} and depth-wise separable and time-dilated convolutions~\cite{drossos:2020:ijcnn}, and we present a variant of the time-dilated convolutions. Our presented method includes a time-dilated convolution that is conditioned on the predictions of the classifier. To evaluate our method, we employ a freely available SED dataset, and we assess the performance on typically used and frame-wise calculated metrics, i.e. $\text{F}_{1}$ score and error rate (ER). The rest of the paper is organized as follows. In Section~\ref{sec:method} we present out method, and we describe the followed evaluation procedure in Section~\ref{sec:evaluation}. The obtained results and their discussion are in Section~\ref{sec:results}, and Section~\ref{sec:conclusions} concludes the paper. 

\section{Proposed method}\label{sec:method}
Our proposed method takes as an input a sequence $\mathbf{X}\in\mathbb{R}^{T\times F}$ of $T$ vectors, with each vector consisting of $F$ features, and applies in series a learnable feature extractor, a temporal pattern identifier, and a classifier. The output of the method is another sequence of $T$ vectors, $\hat{\mathbf{Y}} = [\hat{\mathbf{y}}_{1},\ldots,\hat{\mathbf{y}}_{T}]$, with $\hat{\mathbf{y}}_{t}\in[0,1]^{C}$ representing the activity predictions of $C$ sound event classes for each time step $T$. We employ a learnable feature extractor consisting of DWS convolutions and a time-dilated convolution, used as the temporal pattern identifier. To take advantage of intra- and inter-class activity patterns, we propose a novel algorithm for conditioning the time-dilated convolution, using the predictions of the classifier, like a language modelling. 

\subsection{Learnable feature extractor}
Our learnable feature extractor consists of $L$ DWS convolution blocks. The $l$-th DWS convolution block gets as an input the output of the previous block, $\mathbf{H}_{l-1}\in\mathbb{R}^{H^{\text{c}}_{l-1}\times H^{\text{h}}_{l-1}\times H^{\text{w}}_{l-1}}$, where $H^{\text{c}}_{l-1}$, $H^{\text{h}}_{l-1}$, and $H^{\text{w}}_{l-1}$ are the amount of channels, the height, and width of the output of the $l-1$-th DWS convolution block, respectively. $\mathbf{H}_{0}=\mathbf{X}$ is the input to our method with $H^{\text{c}}_{0}=1$, $H^{\text{h}}_{0}=T$, and $H^{\text{w}}_{0}=F$. The output of the $l$-th DWS block is $\mathbf{H}_{l}\in\mathbb{R}^{H^{\text{c}}_{l}\times H^{\text{h}}_{l}\times H^{\text{w}}_{l}}$. Each DWS convolution block consists of a DWS convolution operation, a normalization process, a sub-sampling process, a non-linearity, and a dropout. 

The DWS convolution operation itself, consists of two convolutions with unit stride, a leaky rectified linear unit (LReLU), and a normalization process. The \emph{first convolution} of the DWS convolution operation of the $l$-th DWS convolution block, learns spatial information on its input $\mathbf{H}_{l-1}$, by utilizing $H^{\text{c}}_{l-1}$ kernels $\mathbf{K}_{l}\in\mathbb{R}^{K^{\text{dh}}_{l}\times K^{\text{dw}}_{l}}$, and is given as
\begin{align}\label{eq:depthwise-conv}
    \mathbf{D}_{l}^{h^{\text{c}}_{l-1},d^{\text{h}}_{l}, d^{\text{w}}_{l}}& = (\mathbf{K}_{l}^{h^{\text{c}}_{l-1}} * \mathbf{H}^{h^{\text{c}}_{l-1}}_{l-1})(h^{\text{h}}_{l-1}-K^{\text{dh}}, h^{\text{w}}_{l-1}-K^{\text{dw}})\nonumber\\
    =\sum\limits_{k^{\text{dh}}=1}^{K^{\text{dh}}}&\sum\limits_{k^{\text{dw}}=1}^{K^{\text{dw}}}\mathbf{H}^{h^{\text{c}}_{l-1}, h^{\text{h}}_{l-1}-k^{\text{dh}}, h^{\text{w}}_{l-1}-k^{\text{dw}}}_{l-1}\mathbf{K}_{l}^{h^{\text{c}}_{l-1}, k^{\text{dh}}, k^{\text{dw}}}\text{,}
\end{align}
\noindent
where $\mathbf{D}_{l}\in\mathbb{R}^{H^{\text{c}}_{l-1}\times D^{\text{h}}_{l}\times D^{\text{w}}_{l}}$ is the output features of the \emph{first convolution} at the $l$-th DWS convolution block. $\mathbf{D}_{l}$ is used as an input to a LReLU and a batch-normalization process as
\begin{align}
    \mathbf{D}'_{l} =& \text{BN}(\text{LReLU}(\mathbf{D}_{l}))\text{, where}\\
    \text{LReLU}(\alpha) =& \begin{cases}
    \alpha, &\text{ if }\alpha\geq0,\\
    \beta \alpha &\text{ otherwise}
    \end{cases}\text{,}
\end{align}
BN is the batch-normalization process, $0\leq\beta<1$ is a hyper-parameter
of the LReLU function, and $\mathbf{D}'_{l}\in\mathbb{R}^{H^{\text{c}}_{l-1}\times D^{\text{h}}_{l}\times D^{\text{w}}_{l}}$ is the output of the BN. The second convolution of the DWS convolution operation learns cross-channel information. It
utilizes $H^{\text{c}}_{l}$ kernels, $\mathbf{z}_{l}^{l}\in\mathbb{R}^{H^{\text{c}}_{l-1}}$, takes as an input $\mathbf{D}'_{l}$, and processes it as
\begin{equation}\label{eq:pointwise-conv}
    \mathbf{S}_{l}^{h^{\text{c}}_{l},d^{\text{h}}_{l}, d^{\text{w}}_{l}} = \sum\limits_{h^{\text{c}}_{l-1}=1}^{H^{\text{c}}_{l-1}}\mathbf{D}'^{h^{\text{c}}_{l-1},d^{\text{h}}_{l},d^{\text{w}}_{l}}_{l}\mathbf{z}_{l}^{h^{\text{c}}_{l},h^{\text{c}}_{l-1}}\text{.}
\end{equation}
\noindent
where $\mathbf{S}_{l}\in\mathbb{R}^{H^{\text{c}}_{l}\times D^{\text{h}}_{l}\times D^{\text{w}}_{l}}$ is the output of of the second convolution operation at the $l$-th DWS convolution operation. The output of the DWS \emph{block} is obtained as
\begin{equation}
    \mathbf{H}_{l}=\text{BN}(\text{ReLU}(\mathbf{S}_{l}))\text{,}
\end{equation}
\noindent
where ReLU is the rectified linear unit. A sub-sampling operation (e.g. max-pooling) over the dimension of features is applied after each BN of a DWS \emph{block}, and a dropout with probability $p$, where $p$ is a hyper-parameter. Appropriate zero-padding is applied by each $\text{DWS}_{l}$ so that the length of the sequence in $\mathbf{H}_{L}$ is the same as the sequence length of $\mathbf{X}$. Finally, $\mathbf{H}_{L}$ is reshaped to $\mathbf{H}'\in\mathbb{R}^{H'^{\text{h}}\times H'^{\text{w}}}$, where $H'^{\text{h}}=H^{h}_{L}$ and $H'^{\text{w}}=H^{c}_{L}\cdot H^{w}_{L}$.

\subsection{Conditioned time-dilated convolutions}
$\mathbf{H}'$ is given as an input to a 2D conditioned time-dilated convolution (CDCNN), which is followed by a linear layer with a softmax activation function that acts as a classifier. A 2D CDCNN is a typical 2D time-dilated convolution (DCNN), which has an extra input channel. This extra input channel is initialized to zeros and, during the forward pass, is populated with jointly-learned embeddings of the predictions of the classifier that follows the DCNN. 
\begin{algorithm}[!t]
\SetAlgoLined
\DontPrintSemicolon
\KwIn{$\mathbf{H}'\in\mathbb{R}^{H'^{\text{h}}\times H'^{\text{w}}}$: input tensor,
$\xi^{h}$: time-dilation factor,
CNN: typical 2D convolution with $K'^{o}$ kernels: $\mathbf{K}'\in\mathbb{R}^{2\times K'^{\text{h}}\times K'^{\text{w}}}$, 
$\mathbf{W}\in\mathbb{R}^{H'^{\text{w}}\times C}$, $\mathbf{b}\in\mathbb{R}^{H'^{\text{w}}}$: weight matrix and bias of the trainable affine transform,
Cls: classifier, $H'^{h}$: sequence dimensionality, $K'^{h}$: number of output feature kernels}
\KwOut{$\hat{\mathbf{Y}}=[\hat{\mathbf{y}}_{1},\ldots,\hat{\mathbf{y}}_{O'_{\text{h}}}]$: Predictions of classifier Cls}
 $O^{\text{h}}\gets H'^{\text{h}} - \xi^{h}K'^{h}$\;
 $\mathbf{Q} \gets [\mathbf{0}]^{H'^{\text{h}}\times H'^{\text{w}}}$\;
 $\hat{\mathbf{Y}} \gets [\mathbf{0}]^{O^{\text{h}}\times C}$\;$\text{cntr}\gets1$\;
 \For{$i \gets \xi^{h}K'^{h}$ \textbf{to} $H'^{\text{h}}$}{
  $\boldsymbol\Psi \gets [\mathbf{H}, \mathbf{Q}]$\tcp*{$\boldsymbol\Psi\in\mathbb{R}^{2\times H'^{\text{h}}\times H'^{\text{w}}}$} 
  $\boldsymbol\Psi' \gets [\boldsymbol\Psi^{:,i-\xi^{h}K'^{h},:}, \boldsymbol\Psi^{:, i-\xi^{h}(K'^{h}-1), :}, \ldots, \boldsymbol\Psi^{:, i, :}]$\;
  $\mathbf{O} \gets \text{CNN}(\boldsymbol\Psi')$\tcp*{$\mathbf{O}\in\mathbb{R}^{K'_{o}\times 1 \times O_{w}}$} 
  $\mathbf{o}' \gets \text{flatten}(\mathbf{O})$\tcp*{$\mathbf{o}'\in\mathbb{R}^{K'_{o}\cdot O_{w}}$}
  $\hat{\mathbf{y}} \gets \text{Cls}(\mathbf{o}')$\tcp*{$\mathbf{y}\in[0, 1]^{C}$}
  $\mathbf{Q}_{i, :} \gets \mathbf{W}\cdot\hat{\mathbf{y}} + \mathbf{b}$\;
  $\hat{\mathbf{Y}}_{\text{cntr},:} \gets \hat{\mathbf{y}}$\;
  cntr $\gets \text{cntr}+1$\;
 }
 \Return{$\hat{\mathbf{Y}}$}\;
 \caption{Conditioned time-dilated convolution}
 \label{alg:dilated}
\end{algorithm}

The 2D DCNN, with $K'^{\text{o}}$ kernels $\mathbf{K}'\in\mathbb{R}^{K'^{\text{h}}\times K'^{\text{w}}}$, operates on $\mathbf{H}'$ as
\begin{align}\label{eq:conv_dilated}
    \mathbf{O}_{k'^{\text{o}}, o^{\text{h}}, o^{\text{w}}}=&(\mathbf{K}'*\mathbf{H}')(h'^{\text{h}}-\xi^{\text{h}} \cdot k'^{\text{h}}, h'^{\text{w}}-\xi^{\text{w}}\cdot k'^{\text{w}})\nonumber\\
    =&\sum\limits_{k'^{\text{h}}=1}^{K'^{\text{h}}}\sum\limits_{k'^{\text{w}}=1}^{K'^{\text{w}}}\mathbf{H}'^{h'^{\text{h}}-\xi^{\text{h}}\cdot k'^{\text{h}}, h'^{\text{w}}-k'^{\text{w}}} \mathbf{K}'^{k'^{\text{o}}, k'^{\text{h}}, k'^{\text{w}}}\text{,}
\end{align}
\noindent
where $\mathbf{O}\in\mathbb{R}^{K'^{\text{o}}\times O^{\text{h}}\times O^{\text{w}}}$ is the output of the 2D DCNN, and $\xi_{\text{h}}\in\mathbb{N}^{\star}$ is the time dilation factor of the kernel $\mathbf{K}'$. The process of a DCNN is illustrated in Figure~\ref{fig:dilated-conv}. 

Then, $\mathbf{O}$ is reshaped to a matrix to represent a sequence of $O_{\text{h}}$ vectors with $O_{\text{w}}\cdot K'_{\text{o}}$ features, i.e. $\mathbf{O}$ is transformed to $\mathbf{O}'\in\mathbf{R}^{O'_{\text{h}}\times(O'_{\text{w}}\cdot K'_{\text{o}})}$. $\mathbf{O}'$ is given as an input to the subsequent classifier $\text{Cls}(\cdot)$ with shared weights through time, to do SED as 
\begin{equation}\label{eq:cls}
    \hat{\mathbf{y}}_{o'_{\text{h}}} = \text{Cls}(\mathbf{o}'_{o'_{\text{h}}})\text{.}
\end{equation}

Taking advantage of the cross-channel learning process of the time-dilated convolutions and inspired by language modelling methods for SED, e.g.~\cite{drossos:2019:dcase}, here we propose a novel algorithm for the conditioning of the time-dilated convolutions, based on Eq.~\eqref{eq:conv_dilated}. With our algorithm, we aim to take advantage of the previous predictions of the classifier and make the time-dilated convolution to learn intra- and inter-class patterns, by processing simultaneously its input features and the predictions of the classifier for multiple time-steps (like a language modelling). To do so, we sequentially calculate each $\mathbf{o}'_{o_{\text{h}}}$. This will allow us to apply Eq.~\eqref{eq:cls} on each $\mathbf{o}'_{o'_{\text{h}}}$, and use the output of the Cls for conditioning the prediction of $\mathbf{o}'_{o'_{\text{h}}+1}$. To overcome any dimensionality mismatching for the conditioning of the prediction of $\mathbf{o}'_{o'_{\text{h}}+1}$, we employ a trainable affine transform with a bias, $\text{Aff}:[0, 1]^{C}\mapsto\mathbb{R}^{H'^{\text{w}}}$. 
Our proposed algorithm is given in Algorithm~\ref{alg:dilated}. The parameters of our proposed method (including the DWS blocks, the Cls, the Aff, and the CDCNN) can be jointly optimized using the typical, for SED, cross-entropy loss and $\hat{\mathbf{Y}}$. 
\begin{figure}[!t]
    \centering
    \subfloat[Processing of $\protect\mathbf{H}'_{h'_{\text{h}},h'_{\text{w}}}$]{%
     \includegraphics[width=.45\columnwidth]{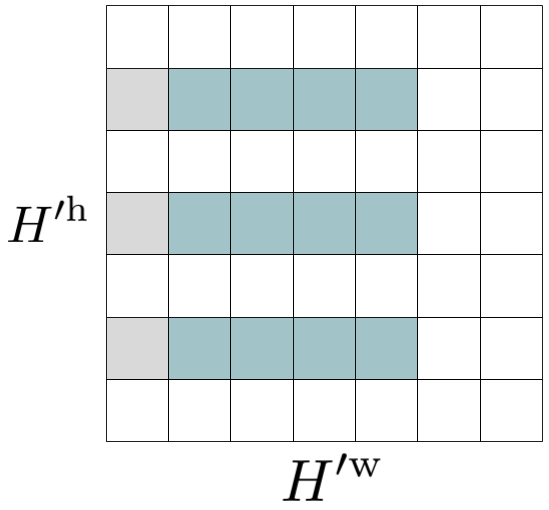}
     }~
    \hfill
     \subfloat[Processing of $\protect\mathbf{H}'_{h'_{\text{h}},h'_{\text{w}}+1}$]{%
       \includegraphics[width=.45\columnwidth]{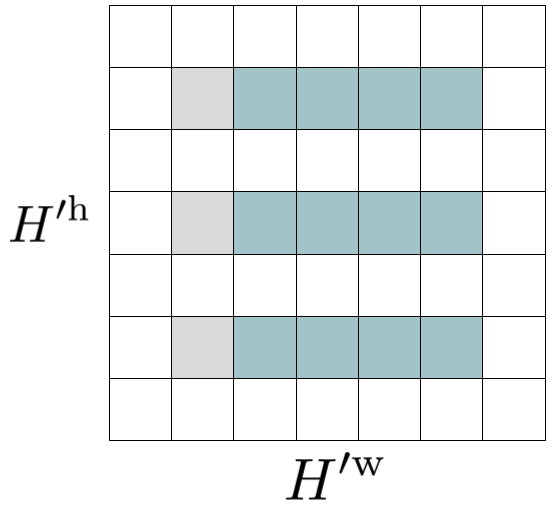}
     }
     \caption{Illustration of the process described in Eq.~\eqref{eq:conv_dilated} using $\xi=2$ and processing two consecutive patches of $\mathbf{H}'$. Squares coloured with cyan signify the elements participating at the processing of $\protect\mathbf{H}'_{h'_{\text{h}},h'_{\text{w}}}$, and coloured with grey are the elements of $\protect\mathbf{H}'_{h'_{\text{h}},h'_{\text{w}}-1}$.}
     \label{fig:dilated-conv}
\end{figure}

\section{Evaluation}\label{sec:evaluation}
To evaluate our method, we employ a publicly available SED dataset, the frame-wise $\text{F}_{1}$ score and error-rate (ER) as the metrics, and a published method with DWS and non-conditioned time-dilated convolutions as our baseline. We choose the specific baseline because it has achieved the best frame-wise $\text{F}_{1}$ and ER values at our employed dataset~\cite{drossos:2020:ijcnn}, and also employs DWS and (non-conditioned) time-dilated convolutions. The code for both our method\footnote{\url{https://github.com/dr-costas/dslam}} and our employed baseline\footnote{\url{https://github.com/dr-costas/dnd-sed}} is based on the PyTorch framework and is available online. 

\subsection{Dataset and data pre-processing}
We employ the TUT-SED Synthetic dataset as our dataset. TUT-SED Synthetic can be found online\footnote{\url{http://www.cs.tut.fi/sgn/arg/taslp2017-crnn-sed/tut-sed-synthetic-2016}} and has been previously used in multiple SED papers~\cite{drossos:2020:ijcnn,cakir:2017:taslp,drossos:2019:dcase,huang:2018:iwaenc}. This dataset consists of 100 mixtures, approximately eight seconds long, and split in 60\%-20\%-20\% fashion for training, validation, and testing (respectively). TUT-SED Synthetic contains overlapping, isolated sound events from $C=16$ classes, namely ``alarms \& sirens'', ``baby crying'', ``bird singing'', ``bus'', ``cat meowing'', ``crowd applause'', ``crowd cheering'', ``dog barking'', ``footsteps'', ``glass smash'', ``gun shot'', ``horse walk'', ``mixer'', ``motorcycle'', ``rain'', and ``thunder'', with a maximum polyphony of 5. From each mixture, we extract non-overlapping sequences of $T=1024$ vectors with $F=40$ log-scaled mel-band energies, using a window of $\approx 22$ ms, with 50\% overlap between successive windows, and the Hamming windowing function. We normalize all extracted feature vectors to have zero-mean and unit-variance, based on statistics calculated on the training split. We use the extracted features $\mathbf{X}$ as the input to our method. 

\subsection{Hyper-parameters, metrics, and evaluation process}
For evaluating the impact of our proposed conditioned time-dilated convolution, we follow the exact same evaluation process with the method employing DWS convolution blocks and non-conditioned time-dilated convolutions~\cite{drossos:2020:ijcnn}. We consider the method using non-conditioned time-dilated convolutions as our baseline (denoted as $\text{Base}$), we employ the same hyper-parameters, and we assess the performance of our proposal using the same metrics as reported for the employed baseline. To have a fair comparison with the Base, we repeat each experiment 10 times and we report the average values of the metrics, over the 10 repetitions of each experiment. In a nutshell, the only variable in the evaluation process affecting the results, is the replacement of the non-conditioned time-dilated convolution with a conditioned one. 

Specifically, we use $L=3$ DWS convolution blocks, with unit stride, $K_{l}^{\text{dh}}=K_{l}^{\text{dw}}=5$, and a padding of $(2, 2)$. The max-pooling operation after each DWS block has kernels and strides of $\{1, 5\}$, $\{1, 4\}$, and $\{1, 2\}$, and for the dropout we use $p=0.25$. Following our baseline, we employ three different kernel sizes for the CDCNN, namely $(K'^{\text{h}}, K'^{\text{w}})\in\{(3, 3), (5, 5), (7, 7)\}$. We employ a dilation factor of $\xi\in\{1, 10, 50, 100\}$ and a zero-padding at the time dimension of $1\cdot\xi$ for the $K'^{\text{h}}= K'^{\text{w}}=3$, $2\cdot\xi$ for $K'^{\text{h}}= K'^{\text{w}}=5$, and $3\cdot\xi$ for $K'^{\text{h}}= K'^{\text{w}}=7$. We use this zero padding in order to have the same time resolution between the input and the output of the CDCNN, i.e. $O'^{\text{h}}=H'^{\text{h}}$. 

We optimize our method on the training split of the TUT-SED Synthetic, employing the Adam optimizer and a batch size of 16. For the optimizer, we employ the values of its hyper-parameters (i.e. $\epsilon$, $\beta_{1}$, and $\beta_{2}$) that are reported at the corresponding paper~\cite{kingma:2015:adam}. For the LReLU we use $\beta=1e-2$, because is the default value of $\beta$ in PyTorch. After each epoch, we measure the loss on the validation split and we stop the optimization process if the validation loss has not improved for 30 consecutive epochs. After the optimization process stops, we utilize the values for the parameters of our method that yielded the lowest loss on the validation split.

For the assessment of the performance of our proposed method, we use the same metrics employed for $\text{Base}$, which are the frame-based $\text{F}_{1}$ score (the higher, the better) and frame-based error rate (ER, the lower the better). We assess all the combinations of $\xi$ and $(K'^{\text{h}}, K'^{\text{w}})$, conducting ten different experiments with each combination, and we report the average and standard deviation of $\text{F}_{1}$ and ER. We denote our method as $\text{CDNN}$ and the different values for $xi$ and $(K'^{\text{h}}, K'^{\text{w}})$ using sub-scripts. For example, $\text{CDNN}_{10, 3}$ is the combination of $\xi=10$ and $K'^{\text{h}}=K'^{\text{w}}=3$. Accordingly, $\text{BASE}_{100, 7}$ denotes the baseline method with the combination of $\xi=100$ and $K'^{\text{h}}=K'^{\text{w}}=7$.

Finally, to evaluate solely the addition of the conditioning without using time-dilated convolution, we also assess the performance of our method with $\xi=1$. That is, we still employ the conditioning as described in Algorithm~\ref{alg:dilated}, but there is no time-dilation for the CDCNN. This case is denoted with a value of 1 for $\xi$, e.g. $\text{CDCNN}_{1, 3}$ for $\xi=1$ and $K'^{\text{h}}=K'^{\text{w}}=3$.

\section{Results and discussion}\label{sec:results}
In Table~\ref{tab:results} are the obtained results for our method, with all the combinations of $\xi$ and $(K'^{\text{h}}, K'^{\text{w}})$. Additionally, in Table~\ref{tab:results} we report the difference of the results obtained with our method that uses the conditioning of the time-dilated convolution, and our baseline which is using a non-conditioned time-dilated convolution. 

\begin{table}[!t]
\centering
\caption{Obtained average (Avg) and STD $\text{F}_{1}$ score and ER for our method, with all the combinations of $\xi$ and $(K'^{\text{h}}, K'^{\text{w}})$. $\Delta_{\text{CDCNN-Base}}$ is the difference between our method (CDCNN) and Base, for the corresponding average value of $\text{F}_{1}$ or ER.}
\smallskip
\label{tab:results}
\resizebox{\columnwidth}{!}{%
\begin{tabular}{lcccc}
 & \multicolumn{2}{c}{$\text{\textbf{F}}_{\boldsymbol1}$} & \multicolumn{2}{c}{\textbf{ER}} \\
 & Avg $\pm$STD & $\Delta_{\text{CDCNN-Base}}$ & Avg $\pm$STD & $\Delta_{\text{CDCNN-Base}}$ \\
 \hline
$\text{CDCNN}_{1,3}$ & 0.59 $\pm$0.01 & 0.00 & 0.54 $\pm$0.01 & 0.00 \\
$\text{CDCNN}_{10,3}$ & 0.63 $\pm$0.02 & 0.01 & 0.49 $\pm$0.02 & -0.03 \\
$\text{CDCNN}_{50,3}$ & 0.61 $\pm$0.02 & 0.00 & 0.52 $\pm$0.02 & -0.01 \\
$\text{CDCNN}_{100,3}$ & 0.60 $\pm$0.02 & 0.00 & 0.54 $\pm$0.02 & 0.01 \\
$\text{CDCNN}_{1,5}$ & 0.61 $\pm$0.01 & 0.02 & 0.52 $\pm$0.01 & -0.01 \\
$\text{CDCNN}_{10,5}$ & 0.64 $\pm$0.01 & 0.02 & 0.49 $\pm$0.02 & -0.03 \\
$\text{CDCNN}_{50,5}$ & 0.61 $\pm$0.01 & -0.01 & 0.52 $\pm$0.01 & 0.00 \\
$\text{CDCNN}_{100,5}$ & 0.58 $\pm$0.01 & 0.00 & 0.56 $\pm$0.01 & 0.00 \\
$\text{CDCNN}_{1,7}$ & 0.61 $\pm$0.02 & 0.01 & 0.51 $\pm$0.02 & -0.03 \\
$\text{CDCNN}_{10,7}$ & \textbf{0.65} $\pm$0.02 & 0.02 & \textbf{0.47} $\pm$0.02 & -0.03 \\
$\text{CDCNN}_{50,7}$ & 0.60 $\pm$0.02 & -0.01 & 0.54 $\pm$0.03 & 0.01 \\
$\text{CDCNN}_{100,7}$ & 0.58 $\pm$0.01 & 0.00 & 0.57 $\pm$0.01 & 0.00
\end{tabular}
}
\end{table}

As can be seen from Table~\ref{tab:results}, using the conditioning of the time-dilated convolution offers a boost mostly when the dilation factor $\xi$ is $1<\xi<50$. This fact reveals that, one hand, the conditioning of the time-dilated convolution benefits the method mostly when there is a time-dilation, i.e., when $\xi>1$. On the other hand, and taking into account the results for $\xi=50$ and $\xi=100$, it can be seen that increased dilation and conditioning, tend to impact the method almost negatively. Though, for $\xi=10$ it can be observed that the method constantly yields the best results for each $(K'^{\text{h}}, K'^{\text{w}})$. 

Finally, in Table~\ref{tab:results} can be observed that the best results are obtained for $\xi=10$ and $\text{CDCNN}_{10, 7}$. This is in accordance with the Base method, where the best results without conditioning of the time-dilated convolutions, are also obtained for $\xi=10$ and $\text{CDCNN}_{10, 7}$. This fact clearly indicates that the conditioning of the time-dilated convolution can improve the performance. Though, more detailed research is needed to explore the learning of the inter- and intra-class activity patterns that the method probably is learning with the conditioning. 

\section{Conclusions}\label{sec:conclusions}
In this paper we presented a novel algorithm for the conditioning of the time-dilated convolutions. Our algorithm mimics the language modelling technique for RNNs, by applying a similar conditioning with the previous predictions of a classifier, but on time-dilated convolutions. To evaluate our method, we employed an existing SED method that is based on DWS and time-dilated convolutions, and we applied our algorithm. We carefully designed the evaluation process, so that the only variable that could affect the results is the application of our algorithm. The obtained results showed that with our algorithm, we SED method could gain a 2\% increase on the $\text{F}_{1}$ score and a 3\% decrease on the ER. 

Future research direction, could be focused on the impact of our proposed algorithm on the learning of inter- and intra-class activity patterns. This could be done, by investigating the performance with our algorithm, on different sound event classes that exhibit these kind of patters, for example ``footsteps'', ``clock ticking'', ``cars passing by''. 
% -------------------------------------------------------------------------
% Either list references using the bibliography style file IEEEtran.bst
\bibliographystyle{IEEEtran}
\bibliography{refs}
\end{sloppy}
\end{document}